\documentclass[conference]{IEEEtran}
\usepackage{graphics}
\usepackage{graphicx}
\usepackage{epsfig,amssymb}
\usepackage{amsmath}
\usepackage{times}
\usepackage{mathrsfs}
\usepackage{array}
\usepackage{amsmath, latexsym, amsfonts, amssymb}
\usepackage[mathscr]{eucal}
\usepackage{array, tabularx}
\usepackage[table]{xcolor}
\usepackage{psfrag}
\usepackage{multirow}
\usepackage{booktabs}
\usepackage{epsfig}
\usepackage{cite}
\usepackage{tikz}

\usepackage{subfigure}
\usepackage{url}

\newcommand{\paren}[1]{\left(#1\right)}
\newcommand{\sqparen}[1]{\left[#1\right]}
\newcommand{\brparen}[1]{\left\{#1\right\}}

\newcommand{\PR}[1]{\ensuremath{\mathsf{Pr}\left\{#1\right\}}} 
\newcommand{\e}[1]{\ensuremath{{\rm e}^{#1}}} 

\newcommand{\sinr}{\ensuremath{{\rm SINR}}}
\newcommand{\snr}{\ensuremath{{\rm SNR}}}

\newcommand{\BO}[1]{\ensuremath{O\paren{#1}}} 
\newcommand{\cnr}[1]{\ensuremath{\mathcal{C N} \left(#1 \right)}} 

\renewcommand{\vec}[1]{\ensuremath{\boldsymbol{#1}}} 
\newcommand{\ie}{\ensuremath{{\text{\em i.e.}}}}

\newtheorem{theorem}{Theorem}
\newtheorem{lemma}{Lemma}
\newtheorem{definition}{Definition}

\IEEEoverridecommandlockouts

\begin{document}
\title{Outage Capacity of Opportunistic Beamforming with Random User Locations}
\author{\IEEEauthorblockN{Tharaka Samarasinghe}
\IEEEauthorblockA{Department of Electrical and \\Computer Systems Engineering,\\ Monash University, Australia.\\
Email: tharaka.samarasinghe@monash.edu}\and \IEEEauthorblockN{Hazer Inaltekin}
\IEEEauthorblockA{Department of Electrical and\\ Electronics Engineering,\\
Antalya International University, Turkey.\\
Email: hazeri@antalya.edu.tr} \and \IEEEauthorblockN{Jamie S.
Evans}
\IEEEauthorblockA{Department of Electrical and \\Computer Systems Engineering,\\ Monash University, Australia.\\
Email: jamie.evans@monash.edu}\vspace{-3cm}
\thanks{This research was supported in part by the European Commission Research Executive Agency Marie Curie FP7-Reintegration-Grants under Grant PCIG10-GA-2011-303713, and in part by the Australian Research Council under Grant DP-11-0102729.}}
\date{}
\bibliographystyle{ieeetr}
\maketitle
\vspace{-0.5cm}
\begin{abstract}
This paper studies the outage capacity of a network consisting of a multitude of heterogenous mobile users, and operating according to the classical opportunistic beamforming framework. The base station is located at the center of the cell, which is modeled as a disk of finite radius. The random user locations are modeled using a homogenous spatial Poisson point process. The received signals are impaired by both fading and location dependent path loss. For this system, we first derive an expression for the beam outage probability. This expression holds for all path loss models that satisfy some mild conditions. Then, we focus on two specific path loss models ({\em i.e.,} an unbounded model and a more realistic bounded one) to illustrate the applications of our results. In the large system limit where the cell radius tends to infinity, the beam outage capacity and its scaling behavior are derived for the selected specific path loss models. It is shown that the beam outage capacity scales logarithmically for the unbounded model. On the other hand, this scaling behavior becomes double logarithmic for the bounded model. Intuitive explanations are provided as to why we observe different scaling behavior for different path loss models. Numerical evaluations are performed to give further insights, and to illustrate the applicability of the outage capacity results even to a cell having a small finite radius.
\end{abstract}

\section{Introduction}
Since its inception in \cite{Tse02}, opportunistic beamforming (OBF) has sparked
a great deal of interest in the wireless communications
research community as an important adaptive signaling
technique that utilizes multiuser diversity and varying channel
conditions to extract full multiplexing gain available in vector
broadcast channels \cite{Tse02, hassibi, Tharaka-TransIT, Bayesteh12, Tharaka-Peva, Huangrao12}. The main advantages of OBF are threefold. It attains the sum-rate capacity with full channel state
information (CSI) to a first order for large numbers of mobile users (MUs) in the
network \cite{hassibi}. Its operation only requires
partial CSI in the form of signal-to-interference-plus-noise
ratios ($\sinr$) leading to a significant reduction in the feedback load. It is an asymptotically feedback optimal transmission strategy \cite{Bayesteh12}. In this paper, we consider the classical opportunistic communication along multiple orthonormal beams in a network consisting of a multitude of heterogenous MUs, and study the outage capacity of the resulting communication system.

In most of the existing work on OBF,  the MUs are assumed to be homogenous and equidistant from the base station (BS) \cite{Tse02, hassibi, Tharaka-TransIT, Bayesteh12}. Recently, works such as \cite{Tharaka-Peva} and \cite{Huangrao12} have focussed on heterogenous networks, which are better representations of practical communication systems where the MUs experience location dependent path loss. In \cite{Tharaka-Peva}, heterogenous MUs are grouped into a finite number of user classes, and the asymptotic throughput scaling of the resulting system is analyzed. In \cite{Huangrao12}, each MU has its own deterministic path loss coefficient, and the authors focus on obtaining an expression for the ergodic capacity.

In this paper, we model the random MU locations using a homogenous spatial Poisson point process (PPP)
of intensity $\lambda$. The signal received by a MU is impaired by both fading and the location dependent path loss. Compared to \cite{Tharaka-Peva} and \cite{Huangrao12}, the path loss coefficients in this paper are random, and governed by a path loss model $G(d)$, where $d$ represents the distance from the BS. 
In this setting, the ergodic capacity achieving transmission strategy involves averaging over all channel variations. 
The requirement to average over location dependent and usually slowly varying path loss values questions the suitability of ergodic capacity as a performance measure for this setup \cite{viswanathtse}. Thus, we focus on the beam outage capacities as a performance metric in this paper, and obtain downlink outage performance of OBF.

Our contributions and the paper organization are as
follows. In Section \ref{Section:System model}, we introduce the system model and formally define the performance measures of interest. The cell is modeled as a disk of radius $D$ with the BS located at the center of the disk. In Section \ref{Section: Max sinr}, we obtain an expression for the beam outage probability for the system in consideration. This expression holds for all path loss models that satisfy some mild conditions. Then, we use this result to derive beam outage probabilities for specific path loss models, and obtain further insights into the downlink outage performance of OBF. We focus on two well known path loss models. Firstly, we study the unbounded power-law path loss model, which has an unrealistic singularity at the origin. Due to the unbounded behavior, the path loss can take any value between zero and infinity in this model. Secondly, we study a more realistic bounded path loss model, where the path loss is always less than one.

In Section \ref{Section: Outage cap and scaling}, we consider the large system limit as $D$ tends to infinity. Using beam outage probability expressions obtained in Section \ref{Section: Max sinr}, we study the outage capacity and its scaling behavior for each of the path loss models. To this end, we obtain expressions that can be easily used to calculate the beam outage capacity of the system of interest. We also show that for the unbounded path loss model, the beam outage capacity behaves according to $\BO{\log\paren{\lambda}}$ as $\lambda$ grows large. On the other hand, for the bounded path loss model, the beam outage capacity behaves according to $\BO{\log\log\paren{\lambda}}$ as $\lambda$ grows large, revealing a different outage capacity scaling behavior. We justify why this difference occurs: It is in fact due to the singularity at the origin in the unbounded path loss model, which makes the $\sinr$ values unbounded.

In Section \ref{Section: Num Eval}, we present some numerical evaluations to provide more insights into our results. To this end, we show that the large system limit closely approximates the beam outage capacity even for cells having a finite radius. In particular, the large system outage capacities are very close to those achieved in cells having a radius of more than one. Moreover, the rate of convergence of these results increases with the MU intensity and the path loss exponent. Section \ref{Section: Conclusions} concludes the paper.

\section{System Model and Problem Setup}\label{Section:System model}

\begin{figure}[t]
\begin{center}
\hspace{0cm}
\begin{tikzpicture}[
        scale=1]
\filldraw[fill=black] (0, 0) circle (3pt);
\filldraw[fill=black] (0, 0.7) circle (1.5pt);
\filldraw[fill=black] (0.2, 2.5) circle (1.5pt);
\filldraw[fill=black] (1, -2) circle (1.5pt);
\filldraw[fill=black] (-0.5, -1.8) circle (1.5pt);
\filldraw[fill=black] (-2, -0.5) circle (1.5pt);
\filldraw[fill=black] (-1.5, 0.1) circle (1.5pt);
\filldraw[fill=black] (-0.2, 0.5) circle (1.5pt);
\filldraw[fill=black] (1.2, 1) circle (1.5pt);
\filldraw[fill=black] (1.6, 0.25) circle (1.5pt);
\filldraw[fill=black] (2.5, 0.1) circle (1.5pt);
\filldraw[fill=black] (0.8, -.6) circle (1.5pt);
\filldraw[fill=black] (0.6, -.8) circle (1.5pt);
\filldraw[fill=black] (-0.4, -1) circle (1.5pt);
\filldraw[fill=black] (-0.6, 2) circle (1.5pt);
\filldraw[fill=black] (-1.1, -2) circle (1.5pt);

\filldraw[fill=black] (-2.6, -.6) circle (1.5pt);
\filldraw[fill=black] (-2.2, 1.8) circle (1.5pt);
\filldraw[fill=black] (-1.8, 1) circle (1.5pt);
\filldraw[fill=black] (2.5, -2) circle (1.5pt);
\filldraw[fill=black] (2.3, 2.5) circle (1.5pt);

\filldraw[fill=black] (0.2, 0.3) circle (1.5pt);
\filldraw[fill=black] (0.5, -0.5) circle (1.5pt);
\draw[thin] (-0.4, -0.5) -- (-0.4, -0.5)node[pos = 0.1, below]{MU $i$};
\filldraw[fill=black] (-0.4, -0.2) circle (1.5pt);
\filldraw[fill=black] (-0.4, 0.4) circle (1.5pt);
\filldraw[fill=black] (0.7, -0.7) circle (1.5pt);

\draw[thin] (0, 0) -- (0, 0)node[pos = 0.1, below]{BS};
\draw (0,0)circle (48pt); \draw[thick, ->] (0,0) -- (1.7, 0);
\draw[thin] (0.8, 0) -- (0.8, 0)node[pos = 0.1, below]{$D$};
  \end{tikzpicture}
\end{center}
\vspace{-0.3cm} \caption{The network model for a particular realization of MU locations.} \label{Fig: Cell}
\end{figure}
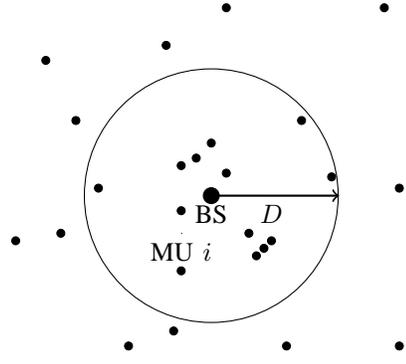

We focus on a single-cell vector broadcast channel. The BS is equipped with $M$ transmitter antennas, and each MU is equipped with a single receive antenna. The cell is
modeled as a disk of radius $D$ 
with the BS located at the center of the disk. MUs are distributed over the plane according
to a PPP of intensity $\lambda$. For a particular realization of MU locations, Fig. \ref{Fig: Cell} gives a
graphical illustration of the part of the plane that includes the cell. Having obtained analytical expressions
for the outage probability in Section \ref{Section: Max sinr}, we will also send $D$ to infinity to obtain outage
capacity expressions in the large system limit in Section \ref{Section: Outage cap and scaling}.

The network operates according to the classical OBF framework as follows. First, the BS generates $M$
random orthonormal beams. Then, it transmits $M$ different symbols,
each of which is drawn from a zero mean and unit variance {\em
circularly-symmetric complex Gaussian} distribution $\cnr{0, 1}$,
in the direction of these beams.  The received signal by a MU is
impaired by both fading and path loss. For MU $i$, it is
given by
\begin{eqnarray}
Y_i =  \sqrt{\rho g_i} \sum_{k=1}^{M}
\vec{h}_i^{\top}\vec{b}_k s_k + Z_i,
\end{eqnarray}
where $\rho$ is the transmit power per beam, $g_i$ is the path
loss coefficient between the $i$th MU and BS, $Z_i$
is the $\cnr{0, 1}$ additive background noise, $\vec{h}_i$ is the
$M$-by-$1$ complex vector containing fading coefficients between the $i$th MU and
BS, $s_k$ and $\vec{b}_k$ are the transmitted symbol and the
beamforming vector corresponding to the $k$th beam, respectively.
We assume that the channel gains are independent and identically
distributed (i.i.d.) random variables drawn from $\cnr{0, 1}$. The
path loss values of all MUs are governed by a path loss model $G(d)$, where $d$ is the distance from the BS.
Therefore, the random path loss values are also i.i.d. among the MUs, where the randomness stems from the fact that MU locations are random. The path loss model is general in the sense that $G$ can be any
function that is continuous, positive, non-increasing, and $G(d)=
\BO{d^{-\alpha}}$ as $d$ grows large for some $\alpha>2$. 

Let $\gamma_{m, i}$ be the $\sinr$ value corresponding to the
$m$th beam at the $i$th MU. Then, it is given by
\begin{eqnarray}
\gamma_{m, i} =  \frac{|\vec{h}_i^{\top} \vec{b}_m|^2}{
\paren{\rho g_i}^{-1} + \sum_{k=1, k\neq m}^{M} |\vec{h}_i^{\top}
\vec{b}_k|^2}. \label{eq:sinr_mb}
\end{eqnarray}
Unlike \cite{hassibi}, for given MU locations, the beam $\sinr$ values are no longer identically distributed among the
MUs, due to the location dependent path loss. Let $F^i_{\gamma}(x)$ represent the cumulative distribution function (CDF) of the beam $\sinr$ at MU $i$.
Using techniques similar to those used in \cite{hassibi}, it is not hard to show that $F_{\gamma}^{i}(x)$, for a given path loss value $g_i = g$, is written as
\begin{eqnarray}
F^i_{\gamma}(x|g_i=g)=1-\frac{e^{-\frac{x}{g \rho
}}}{(x+1)^{M-1}}\label{eq:FX sinr given g}
\end{eqnarray}
for all $i$. Since the MU locations are modeled using a PPP, the CDF of the
distance of a MU from the BS can be written as
$F_D(d)=\paren{\frac{d}{D}}^2$. Also, since $G$ is non-increasing, the CDF of the path
loss of a MU can be written as
\begin{eqnarray}
F_G(g)=1-F_D(G^{-1}(g))=1-\sqparen{\frac{G^{-1}(g)}{D}}^2.\label{eq:FX G}
\end{eqnarray}
Here, we define $G^{-1}(g)$ as $G^{-1}(g) = \inf \brparen{d: G(d) \leq g}$. We note that this definition allows jump discontinuities in $F_G(g)$. An example of such a path loss model is given in Section \ref{Section: Num Eval}.

Each MU feeds back its $\sinr$ information to the BS, and the BS
selects the MU with the highest $\sinr$ on each beam to maximize
the communication rate\footnote{In this paper, we only focus on rate maximization in the network. Interested readers are referred to \cite{Tse02,Tharaka-Peva,Huangrao12} for techniques that can be used to achieve fairness in such a network.}. Therefore, given there are $N$ MUs in
the cell, the instantaneous rate on beam $m$ (measured in terms of nats/s/Hz) can be written as
\begin{eqnarray}
r_m =  \log\paren{1 + \max_{0\leq i \leq N} \gamma_{m,i}}.
\label{Eqn: Rate}
\end{eqnarray}
We say that an outage event occurs on beam $m$ when $r_m$ is less than a target rate value $x$. Thus, the beam outage
probability, or the CDF of the rate on a beam, can be calculated as
\begin{equation}
F_r(x)=  \PR{\log\paren{1 + \max_{0\leq i \leq N}
\gamma_{m,i}} \leq x }   =
F^\star_{\gamma}\paren{\e{x}-1}, \label{Eqn: Rate outage}
\end{equation}
where $F^\star_{\gamma}$ is the CDF of the maximum $\sinr$
on a beam. 

We are interested on the beam outage capacity, which we formally define as follows.
\begin{definition}\label{def: outage cap}
The beam outage capacity  $C_{\rm out}\paren{\epsilon}$ is defined as the supremum of communication rates on a beam that results in
an outage probability of less than $\epsilon$ for that particular beam, $\ie$,
\begin{eqnarray}
C_{\rm out}\paren{\epsilon} = \sup \brparen{x: F_r\paren{x} \leq \epsilon},
\end{eqnarray}
for $\epsilon \in (0,1)$.
\end{definition}
Using this definition, and the monotonicity of $F_{\gamma}^\star$, the beam outage capacity can be written as
\begin{eqnarray}
C_{\rm out}\paren{\epsilon} = \log\paren{F^{\star^{-1}}_{\gamma}\paren{\epsilon}+1},
\label{Eqn: outage capacity}
\end{eqnarray}
where ${F_{\gamma}^\star}^{-1}\paren{\epsilon} = \inf \brparen{x: F_{\gamma}^\star(x) \geq \epsilon}$.
In the next section, we will focus on obtaining expressions for $F_r$ and $F^\star_{\gamma}$, which will be, in turn, used to derive
beam outage capacity expressions for specific path loss models in Section \ref{Section: Outage cap and scaling}.

\section{Beam Outage Probability for the General Path Loss Model}\label{Section: Max sinr}

In our set-up, the number of MUs in the cell is a Poisson distributed random variable with mean $\lambda \pi D^2$.
Hence, in order to derive beam outage probabilities, we will first condition on $N$ and path loss values, and then we will remove
conditioning by averaging over the location process. These ideas are formally presented in the following theorem.
\begin{theorem}\label{Thm: max SINR}
For a given communication rate $x$, the beam outage probability $F_r(x)$ is equal to $F_{\gamma}^\star\paren{\e{x} - 1}$,
where $F_{\gamma}^\star\paren{x}$ is given by
\begin{eqnarray}
F^\star_{\gamma}(x)= \exp{\paren{\frac{-\lambda \pi}{\paren{x+1}^{M-1}}\int_{0}^{D^2}\exp\paren{\frac{-x}{G(\sqrt{t})\rho}}dt}}. \label{Eqn: CDF maxSINR}
\end{eqnarray}
\end{theorem}
\begin{IEEEproof}
Conditioning on $N$ and $\vec{g} = \paren{g_1, \ldots, g_N}^\top$, we have
\begin{eqnarray*}
F^\star_{\gamma}(x|N,\vec{g})= \prod^{N}_{i=1}F^i_{\gamma}(x|g_i),
\end{eqnarray*}
where $\vec{g}$ is the vector containing the path loss values of all the MUs in the cell. 
Averaging over the i.i.d. path loss values gives us
\begin{eqnarray*}
F^\star_{\gamma}(x|N)= \paren{\int_{G(D)}^{G(0)}F_{\gamma}(x|v)d F_G(v)}^N.
\end{eqnarray*}
Similarly, by observing that $\PR{N = n} = \frac{\e{-\lambda \pi D^2} \paren{\lambda \pi D^2}^n}{n!}$, we can uncondition on the number of MUs, and obtain
\begin{equation*}
F^\star_{\gamma}(x)= \sum_{n=0}^{\infty}\frac{\e{-\lambda\pi D^2}\paren{\lambda\pi D^2}^n}{n!}\paren{\int_{G(D)}^{G(0)}F_{\gamma}(x|v)d F_G(v)}^n.
\end{equation*}
%
%
%

Now, by using \eqref{eq:FX sinr given g}, we have
\begin{multline*}
F^\star_{\gamma}(x)= \sum_{n=0}^{\infty}\frac{\e{-\lambda\pi D^2}\paren{\lambda\pi D^2}^n}{n!} \\ \times \paren{\int_{G(D)}^{G(0)}d F_G(v)- \int_{G(D)}^{G(0)}\frac{\e{\frac{-x}{v\rho}}}{\paren{x+1}^{M-1}}d F_G(v) }^n.
\end{multline*}
Since $\int_{G(D)}^{G(0)}d F_G(v)=1$, we get
\begin{eqnarray*}
F^\star_{\gamma}(x)
= \exp{\paren{-\lambda \pi D^2 \int_{G(D)}^{G(0)}\frac{\e{\frac{-x}{v\rho}}}{\paren{x+1}^{M-1}}d F_G(v)}}
\end{eqnarray*}
by writing the infinite summation using the exponential function.
Substituting for $F_G(v)$ from \eqref{eq:FX G} and making a variable change $\paren{G^{-1}(v)}^2=t$ completes the proof.
\end{IEEEproof}

For a given communication rate $x$, the beam outage probability can be obtained easily using \eqref{Eqn: CDF maxSINR}. However, analyzing the outage capacity using this
expression is not straightforward due to the integral that depends on the path loss model. Therefore, in the next subsection,
we apply this result to derive beam outage probabilities for specific path loss models, providing us with further insights.

\subsection{Beam Outage Probabilities for Specific Path Loss Models}\label{Subsection: Applications}

First, we focus on the classical unbounded path loss model, which is $G(d)= d^{-\alpha}$, where $\alpha>2$, {\em e.g.}, see \cite{Sousa90,Sousa92,Andrews11}.
The following lemma gives us the beam outage probability expression for this case.
\begin{lemma}\label{Lemma: max SINR unbound}
Let $G(d)= d^{-\alpha}$, where $\alpha>2$. For a given communication rate $x$, the beam outage probability $F_{{\rm ub}, r}(x)$ is equal
to $F_{{\rm ub},\gamma}^\star\paren{\e{x} - 1}$,
where $F_{{\rm ub},\gamma}^\star\paren{x}$ is given by
\begin{equation}
F^\star_{{\rm ub}, \gamma}(x)= \exp{\paren{\frac{-2\lambda \pi}{\alpha \paren{x+1}^{M-1}}\paren{\frac{\rho}{x}}^{\frac{2}{\alpha}}\gamma\paren{\frac{2}{\alpha},\frac{x D^\alpha}{\rho}} }}, \label{Eqn: CDF maxSINR unbound}
\end{equation}
and $\gamma(\cdot)$ is the lower incomplete gamma function.
\end{lemma}
\begin{IEEEproof}
From Theorem \ref{Thm: max SINR}, we have
\begin{eqnarray*}
F^\star_{{\rm ub}, \gamma}(x)= \exp{\paren{\frac{-\lambda \pi}{\paren{x+1}^{M-1}}\int_{0}^{D^2}\exp\paren{\frac{-x t^{\frac{\alpha}{2}}}{\rho}}dt}},
\end{eqnarray*}
and evaluating the integral completes the proof \cite{ryzhik}.
\end{IEEEproof}

The above path loss model has been extensively used in the literature due to its mathematical tractability. However,
this model has an unrealistic singularity at the origin, which might lead to flawed conclusions \cite{Inaltekin09}. 
Therefore, we also obtain the beam outage probability for a more realistic bounded gain path loss model.
To this end, we choose $G(d)$ as $G(d) = \paren{1 + d^\alpha}^{-1}$, where $\alpha > 2$, {\em e.g.,} see \cite{Inaltekin09,Tharaka-Ausctw2013,Tharaka-VTC2013}.
\begin{lemma}\label{Lemma: max SINR bound}
Let $G(d)= (1+d^{\alpha})^{-1}$, where $\alpha>2$. For a given communication rate $x$, the beam outage probability $F_{{\rm b}, r}(x)$ is equal
to $F_{{\rm b},\gamma}^\star\paren{\e{x} - 1}$,
where $F_{{\rm b},\gamma}^\star\paren{x}$ is given by
\begin{eqnarray}
F^\star_{{\rm b}, \gamma}(x)= \exp{\paren{\frac{-2\lambda \pi \e{\frac{-x}{\rho}}}{\alpha \paren{x+1}^{M-1}}\paren{\frac{\rho}{x}}^{\frac{2}{\alpha}}\gamma\paren{\frac{2}{\alpha},\frac{x D^\alpha}{\rho}} }}, \label{Eqn: CDF maxSINR bound}
\end{eqnarray}
and $\gamma(\cdot)$ is the lower incomplete gamma function.
\end{lemma}

Since the proof follows from the same lines of the proof of Lemma \ref{Lemma: max SINR unbound}, we skip it to avoid repetition.Using above derived expressions for beam outage probabilities, we will analyze the beam outage capacity and its scaling behavior in the next section.

\section{Beam Outage capacity and Its Scaling Behavior}\label{Section: Outage cap and scaling}

In the remaining part of the paper, we will focus on the large system limit as $D$ tends to infinity. When $D$ grows large, the
lower incomplete gamma functions in \eqref{Eqn: CDF maxSINR unbound} and \eqref{Eqn: CDF maxSINR bound} can be approximated by the
gamma function $\Gamma(\cdot)$, $\ie$, we get 
\begin{eqnarray}
F^\star_{{\rm ub}, \gamma}(x)= \exp{\paren{\frac{-2\lambda \pi}{\alpha \paren{x+1}^{M-1}}\paren{\frac{\rho}{x}}^{\frac{2}{\alpha}}\Gamma\paren{\frac{2}{\alpha}} }} \label{Eqn: CDF maxSINR unbound dinf}
\end{eqnarray}
and
\begin{eqnarray}
F^\star_{{\rm b}, \gamma}(x)= \exp{\paren{\frac{-2\lambda \pi \e{\frac{-x}{\rho}}}{\alpha \paren{x+1}^{M-1}}\paren{\frac{\rho}{x}}^{\frac{2}{\alpha}}\Gamma\paren{\frac{2}{\alpha}} }}. \label{Eqn: CDF maxSINR bound dinf}
\end{eqnarray}
We will first obtain beam outage capacity and its scaling behavior for the unbounded path loss model through the following theorem.
\begin{theorem}\label{Thm: outage capacity unbound}
Let $y^\star$ be the solution of
\begin{eqnarray}
y^{a+1} - y^{a} - \paren{\frac{-b}{\log \epsilon}}^{\frac{\alpha}{2}}=0, \label{Eqn: cap sol unbounded}
\end{eqnarray}
where $a=\frac{\alpha}{2}\paren{M-1}$,  $b=\frac{2\lambda\pi}{\alpha}\Gamma\paren{\frac{2}{\alpha}}\rho^{\frac{2}{\alpha}}$ and $y \in \paren{1,\infty}$.
Then, for $G(d) = d^{-\alpha}$, $\alpha > 2$, the beam outage capacity $C_{\rm out, ub}\paren{\epsilon}$ in the large system limit is equal to
$\log\paren{y^\star}$. Moreover, $C_{\rm out, ub}\paren{\epsilon}$ scales according to $\BO{\log\paren{\lambda}}$ as the MU intensity $\lambda$ grows large.
\end{theorem}
\begin{IEEEproof}
We will only focus on the $M>1$ case. $M=1$ case follows from the same lines. From \eqref{Eqn: Rate outage}, \eqref{Eqn: outage capacity} and \eqref{Eqn: CDF maxSINR unbound dinf}, the beam outage capacity $C_{\rm out, ub}\paren{\epsilon}$ should satisfy
\begin{eqnarray*}
\paren{\frac{-b}{\log \epsilon} }^{\frac{\alpha}{2}}= \e{a C_{\rm out, ub}\paren{\epsilon}}\paren{\e{C_{\rm out, ub}\paren{\epsilon}}-1}
\end{eqnarray*}
as $D \rightarrow \infty$. Setting $\e{C_{\rm out, ub}\paren{\epsilon}}=y$ gives us \eqref{Eqn: cap sol unbounded}.
It is not hard to show that $y^{a+1} - y^{a} - \paren{\frac{-b}{\log \epsilon}}^\frac{\alpha}{2}$ is a strictly increasing function of $y$ that tends to infinity as $y$ grows large,
and is negative as $y$ approaches to one. Therefore, $y^\star$ is unique, and its logarithm gives the beam outage capacity of the system without any ambiguity.

Also, from \eqref{Eqn: cap sol unbounded},
\begin{eqnarray*}
\log{y^\star}= \frac{\alpha}{2 a} \log \lambda - \frac{1}{a}\log{\paren{y^\star-1}} + \BO{1}.
\end{eqnarray*}
$\log{y^\star}$ scales according to $\BO{\log \paren{\lambda}}$, which implies $C_{\rm out, ub}\paren{\epsilon}$ scales according to $\BO{\log \paren{\lambda}}$ as $\lambda$ grows large.
\end{IEEEproof}

According to Theorem \ref{Thm: outage capacity unbound}, we can obtain the beam outage capacity by using a root finding algorithm to find the unique $y^\star$
solving \eqref{Eqn: cap sol unbounded} for any value of $M$. Further, when $M=1$, we can get a closed form expression for the beam outage capacity as
\begin{eqnarray*}
C_{\rm out, ub}\paren{\epsilon}= \log\paren{1+\rho\paren{\frac{-2 \lambda \pi  \Gamma\paren{\frac{2}{\alpha}}}{\alpha \log \epsilon}}^\frac{\alpha}{2}}.
\end{eqnarray*}
The beam outage capacity expression above for $M=1$ clearly indicates the logarithmic outage capacity scaling with $\lambda$.

Next, we will obtain similar results for the bounded gain path loss model.
\begin{theorem}\label{Thm: outage capacity bound}
Let $y^\star$ be the solution of
\begin{eqnarray}
\log{\paren{y^a\paren{y-1}}}+ \frac{\alpha}{2 \rho} \paren{y-1}- \frac{\alpha}{2}\log{\paren{\frac{-b}{\log \epsilon}}}=0, \label{Eqn: cap sol bounded}
\end{eqnarray}
where $a=\frac{\alpha}{2}\paren{M-1}$,  $b=\frac{2\lambda\pi}{\alpha}\Gamma\paren{\frac{2}{\alpha}}\rho^{\frac{2}{\alpha}}$ and $y \in \paren{1,\infty}$.
Then, for $G(d) = (1+d^{\alpha})^{-1}$, $\alpha > 2$, the beam outage capacity $C_{\rm out, b}\paren{\epsilon}$ in the large system limit is equal to
$\log\paren{y^\star}$. Moreover, $C_{\rm out, b}\paren{\epsilon}$ scales according to $\BO{\log\log\paren{\lambda}}$ as the MU intensity $\lambda$ grows large.
\end{theorem}
\begin{IEEEproof}
From \eqref{Eqn: Rate outage}, \eqref{Eqn: outage capacity} and \eqref{Eqn: CDF maxSINR bound dinf}, the beam outage capacity $C_{\rm out, b}\paren{\epsilon}$ should satisfy
\begin{equation*}
\paren{\frac{-b}{\log \epsilon} }^{\frac{\alpha}{2}}= \e{a C_{\rm out, b}\paren{\epsilon} + \frac{\alpha}{2 \rho}\paren{\e{C_{\rm out, b}\paren{\epsilon}}-1}}\paren{\e{C_{\rm out, b}\paren{\epsilon}}-1}
\end{equation*}
as $D \rightarrow \infty$. Taking logarithm of both sides and setting $\e{C_{\rm out, b}\paren{\epsilon}}=y$ give us \eqref{Eqn: cap sol bounded}.
It is not hard to show that this is a strictly increasing function of $y$ that tends to infinity as $y$ grows large, and is negative as $y$ approaches to one. Therefore, $y^\star$ is unique, and its logarithm gives
the beam outage capacity of the system without any ambiguity.

Also, from \eqref{Eqn: cap sol bounded},
\begin{eqnarray*}
y^\star&=& \rho \log \lambda - \frac{2 \rho a}{\alpha}\log{\paren{y^\star}} - \frac{2 \rho}{\alpha}\log{\paren{y^\star-1}}+ \BO{1}\\
&=& \rho \log \lambda + \BO{\log\log \lambda}.
\end{eqnarray*}
Therefore, $y^\star$ scales according to $\BO{\log \paren{\lambda}}$, which implies that $C_{\rm out, b}\paren{\epsilon}$ scales according to $\BO{\log\log \paren{\lambda}}$ as $\lambda$ grows large.
\end{IEEEproof}

Theorem \ref{Thm: outage capacity bound} reveals a different outage capacity scaling behavior than that of Theorem \ref{Thm: outage capacity unbound}.
This difference in scaling is in fact caused by the singularity at the origin in the unbounded path loss model.
When $M=1$, it is easy to see that the $\snr$ values become unbounded in the unbounded path loss model, which in turn leads to different scaling behaviors.
For $M>1$, we are almost guaranteed to have at least one MU in the small vicinity $\delta$ of the BS such that its inter-beam interference is practically nulled out, for large values of $\lambda$ as a function of $\delta$ ($\ie$, opportunistic nulling). For such a user, there is only power gain coming from the fading process in the bounded case. On the other hand, in the unbounded path loss model, there is also an extra power gain coming from the singularity at the origin, which results in different scaling behaviors.
%
Similar to the unbounded case, we can use a common root finding algorithm on \eqref{Eqn: cap sol bounded} to
find the beam outage capacity when $G(d)=(1+d^\alpha)^{-1}$.

In the next section, we will present some numerical evaluations to provide further insights into our results.

\section{Numerical Evaluations} \label{Section: Num Eval}

\begin{figure}[t]
\centering{\includegraphics[scale=0.45]{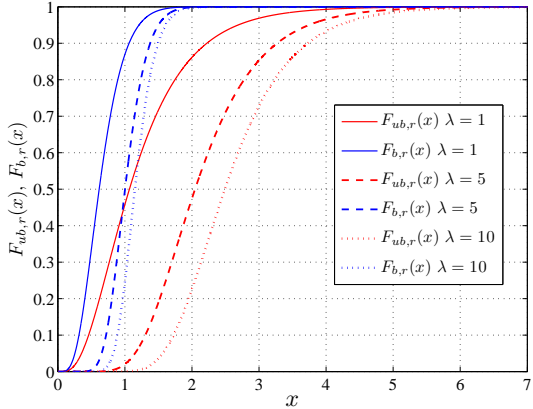}} \caption{Graphical illustration of beam outage probabilities for different values of $\lambda$, where $\rho=1$, $M=2$ and $\alpha=4$.}\vspace{-0.3cm} \label{Fig: CDF}
\end{figure}

We will start by giving a graphical illustration of beam outage probabilities as a function of target communication rates $x$ for each of the path loss models in Fig. \ref{Fig: CDF}. 
We can see that the unbounded model achieves a better beam outage probability with a larger dynamic range because the path loss gain in this case can take any value between zero and infinity. Also, the beam outage probability curves shift right with increasing values of $\lambda$, illustrating the multiuser diversity
gains analyzed in Theorems \ref{Thm: outage capacity unbound} and \ref{Thm: outage capacity bound}.

\begin{figure}[t]
\centering{\includegraphics[scale=0.45]{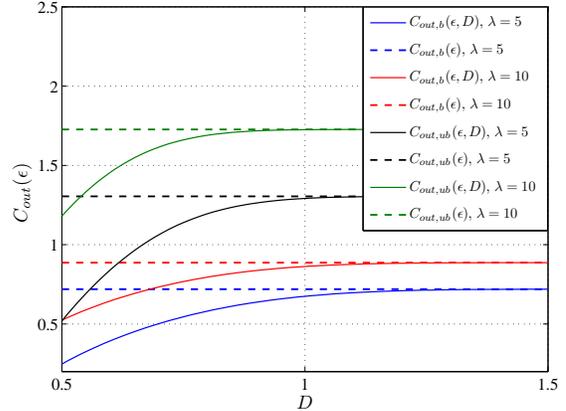}} \caption{The behavior of the beam outage capacity with the cell radius for different values of $\lambda$, where $\rho=1$, $\epsilon=0.1$, $M=2$, and $\alpha=4$.}\vspace{-0.3cm} \label{Fig: Cap with D}
\end{figure}

In Section \ref{Section: Outage cap and scaling}, we have focused on the large system limit as $D$ tends to infinity.
Therefore, the beam outage capacity results
in Theorems \ref{Thm: outage capacity unbound} and \ref{Thm: outage capacity bound} are true for a cell of infinite radius.
However, by using the results in Lemmas \ref{Lemma: max SINR unbound} and \ref{Lemma: max SINR bound}, we can also numerically evaluate the beam outage
capacity for a cell of finite radius. The probability of a MU at the cell edge having the maximum $\sinr$ decreases
with the cell radius, due to the path loss. Therefore, intuitively, the beam outage capacity results in the large system limit should closely approximate the finite case after some value of $D$. To this end, Fig. \ref{Fig: Cap with D} illustrates the behavior of the beam outage capacity as a function of the cell radius. In this figure, $C_{\rm out, ub}(\epsilon, D)$ and $C_{\rm out, b}(\epsilon, D)$ represent the beam outage capacities of the two path loss models for finite $D$. It shows that the beam outage capacity in the large system limit closely approximates the beam outage capacities even for small finite values of $D$. $D$ does not need to be very large
for the results to match, and the large system beam outage capacities are very close to those achieved in cells having a radius of more than one.

\begin{figure}[t]
\centering{\includegraphics[scale=0.45]{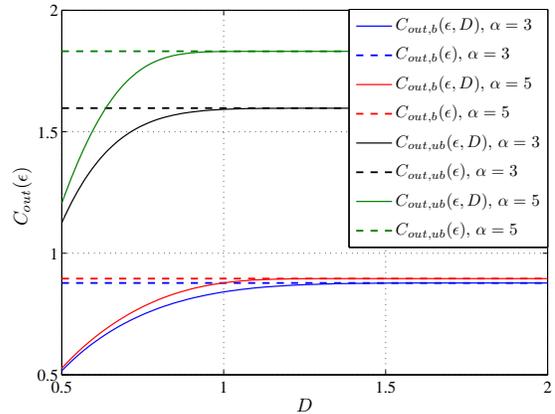}} \caption{The behavior of the beam outage capacity with the cell radius for different values of $\alpha$, where $\rho=1$, $\epsilon=0.1$, $M=2$, and $\lambda=10$.}\vspace{-0.3cm} \label{Fig: Cap with D alpha}
\end{figure}

Furthermore, we can observe that the
convergence is faster for the unbounded path loss model in Fig. \ref{Fig: Cap with D}. In both models, there is a high probability of a MU staying close to the
BS being scheduled for communication. However, this probability is comparatively higher in the unbounded model due to the
unbounded gain. Therefore, its dependence on the cell radius is less prominent compared to the bounded one, allowing the faster convergence.
Secondly, a faster convergence can be observed with increasing values of $\lambda$ as well. This is because increasing $\lambda$ increases the number
of MUs per unit area, which includes the number of MUs staying close to the BS. This again makes the cell edge MUs less prominent, causing faster convergence. By observing \eqref{Eqn: CDF maxSINR unbound} and \eqref{Eqn: CDF maxSINR bound}, we can expect the rate of convergence to increase with $\alpha$ as well. This is illustrated in Fig. \ref{Fig: Cap with D alpha}. This result is expected intuitively as well because increasing $\alpha$ increases the path loss at cell edge MUs, making them less prominent.

\begin{figure}[t]
\centering{\includegraphics[scale=0.45]{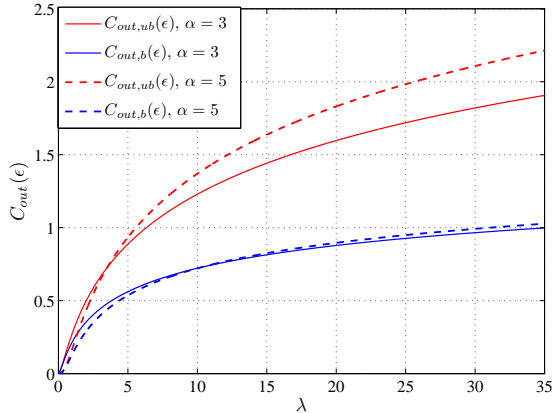}} \vspace{-0.2cm}\caption{The behavior of the beam outage capacity with the MU intensity, where $\rho=1$, $\epsilon=0.01$ and $M=2$.}\vspace{-0.3cm} \label{Fig: outwithlam}
\end{figure}

Finally, we illustrate how the beam outage capacity changes with $\lambda$ in Fig. \ref{Fig: outwithlam}. 
We can clearly observe the $\BO{\log \paren{\lambda}}$ scaling behavior for the unbounded model, and
$\BO{\log\log \paren{\lambda}}$ scaling behavior for the bounded model, which are in line with the results in Theorems \ref{Thm: outage capacity unbound}
and \ref{Thm: outage capacity bound}. It is interesting to note that when $\lambda$ is relatively small, the beam outage capacity first decreases with
$\alpha$, and then increases with $\alpha$ when $\lambda$ is large. The decrease with $\alpha$ is rather intuitive because increasing $\alpha$ increases
the path loss, which decreases the $\sinr$ and the rate. However, note that when $d<1$, $G(d)$ increases with $\alpha$.
As mentioned earlier, when we increase $\lambda$, more prominence is given to the MUs staying close to the BS, $\ie$, to the MUs having distance
less than one. Therefore, at high values of $\lambda$, the beam outage capacity increases with $\alpha$. This somewhat counter intuitive behavior is especially more pronounced for the unbounded path loss model. To overcome it in the bounded case, one can use a path
loss model taking the form of $G(d) = \max(d_0, d)^{-\alpha}$, where $d_0$ is a constant that accounts for a
guard zone around the BS up to a certain distance. $G(d) = \paren{1+ d}^{-\alpha}$ is another option. Due to the generality of the path loss model definition in
Section \ref{Section:System model}, and the result obtained in Theorem \ref{Thm: max SINR}, our analysis can be easily extended to both of these path loss models.

\section{Conclusions}\label{Section: Conclusions}
In this paper, we have studied the outage capacity of a network consisting of a multitude of mobile users whose random locations are modeled using a homogenous spatial Poisson point process, and operating according to the classical opportunistic beamforming framework. Considering a cell modeled as a disk of radius $D$, we have first obtained an expression for the beam outage probability. This expression holds for all path loss models that satisfy some mild conditions. Then, we have applied this result to two well known path loss models. Firstly, we have considered the classical unbounded path loss model, which is $G(d)= d^{-\alpha}$, where $\alpha>2$ and $d$ represents the distance from the base station. Secondly, we have considered a more realistic bounded gain path loss model $G(d) = \paren{1 + d^\alpha}^{-1}$, where $\alpha > 2$. Then, in a large system setting where $D$ tends to infinity, we have obtained analytical expressions for the beam outage capacity and its scaling behavior for each of these path loss models. In particular, we have obtained expressions that can be easily used to calculate the beam outage capacity of the system of interest. We have shown that the beam outage capacity behaves according to $\BO{\log\paren{\lambda}}$ for the unbounded model, and according to $\BO{\log\log\paren{\lambda}}$ for the bounded model, as the user intensity $\lambda$ grows large. The difference in outage capacity scaling is due to the unrealistic singularity in the unbounded model at $d=0$. We have also performed numerical evaluations to give further insights into the derived analytical results describing the network performance. To this end, we have shown that the large system limit closely approximates the beam outage capacity even for cells having a finite radius. In particular, the large system outage capacities are very close to those achieved in cells having a radius of more than one.

\bibliography{Tharaka-bibfile-SG}
\end{document}